\documentstyle[aps,prd]{revtex}
\def\Msol{\rm M_{\odot}}
\def\Tsol{\rm T_{\odot}}

\begin{document}

\draft
\title{Detection of Variable Frequency Signals
Using a Fast Chirp Transform}
\author{F. A. Jenet\\ T. A. Prince}
\address{Division of Physics, Mathematics, and Astronomy and the
LIGO Laboratory,\\ California Institute of Technology,\\
Pasadena, CA 91125}
\date{\today}
\maketitle
\begin{abstract}
The detection of signals with varying frequency is important in many
areas of physics and astrophysics. The current work was motivated
by a desire to detect gravitational waves from
the binary inspiral of neutron stars and black holes,
a topic of significant interest for the  new generation of interferometric
gravitational wave detectors such as LIGO. However, this work has
significant generality beyond gravitational wave signal detection.

We define a Fast Chirp Transform
(FCT) analogous to the Fast Fourier Transform (FFT).  Use of the FCT provides
a simple and powerful formalism for detection of signals with
variable frequency just as Fourier transform techniques
provide a formalism for the detection of signals of constant
frequency.  In particular, use of the FCT can alleviate the
requirement of generating complicated families of filter
functions typically required in the conventional matched filtering process.
We briefly discuss the
application of the FCT to several signal detection problems of current
interest.
\end{abstract}

\section{Introduction}

The detection of periodic signals is a well-developed art.  In contrast,
the detection of signals with variable frequency is an
active area of research in signal processing. Considerable progress
has been made in recent years using a variety of time-frequency
techniques which include wavelets, bilinear transforms, and Short Time
Fourier Transforms (STFTs) \cite{lo88,cht98}.

In this paper, we consider the detection of deterministic signals with
unknown parameters. The case of deterministic signals with unknown
amplitude, phase, frequency, and arrival time has been treated in the
literature \cite{hel68,kay93}. In this paper, we generalize to
an arbitrary number of parameters and consider signals with a
deterministic, but parameterized, frequency evolution.  We call these
``variable frequency signals''. Specifically, we consider signals of
the form:

\begin{equation}
h_s (t) = \left\{ \begin{array}{ll}
	A(t) \text{cos} (\phi(t,\lambda_0,...,\lambda_M)) & 0 < t < T \\
	0 & \text{otherwise}
	\end{array}
	\right.
\label{eq:phasefunc}
\end{equation}

\noindent
where $\phi(t,\lambda_0,...,\lambda_M)$ is real and the
$\{\lambda_0,...,\lambda_M\}$ represent various parameters which
describe the phase evolution (e.g. frequency, frequency derivative,
etc.). In general, $\phi()$ may depend non-linearly on time. For this
paper, the ``instantaneous frequency'' of $h_s$ must be a well-defined
quantity and the frequency evolution of the signal must be
well-resolved in the data. Variable frequency signals include the
class of signals usually known as chirps, i.e. signals which have a
monotonically increasing or decreasing instantaneous frequency.  Such
signals appear in many contexts, such as almost-periodic signals with
a small frequency drift or periodic signals emitted from accelerated
systems.  Chirps are discussed often in the literature of signal
processing \cite{jay87,kay93,cht98}. We will concentrate on chirp
signals in order to simplify the description of the chirp transform
algorithm.

A standard technique the for detection of signals in the presence of
noise is the ``matched filter'' technique \cite{hel68,van68}.
Detection of a signal $h_s(t)$ in the presence of white noise in a
data stream $h(t)$ of length $T$ is based on the matched filter
output:

\begin{equation}
{\cal S} ~=~  \int_{0}^{T} dt~ h_s(t) h(t)
~=~ \int_{0}^{T} d\tau~ h_{filt}(\tau) h(T-\tau)
\label{eq:timematch}
\end{equation}

\noindent
where $h_{filt}(t)$ is an optimal filter function in the time domain.
For white noise, the optimum filter has an impulse response given by
$h_{filt}(\tau )=h_s(T-\tau)$, in the interval $0<\tau<T$.
To detect a signal beginning at time $t_0$ in a data stream of arbitrary length
we compute the quantity:

\begin{equation}
{\cal S}(t_0) ~=~ \int_{-\infty}^{\infty} d\tau~ h_{filt}(\tau) h((t_0+T)-\tau)
\end{equation}

\noindent
where $h(t)$ is zero outside a finite region of interest.
When colored noise is present, it is conventional to work in the frequency
domain. For
matched filtering  of real signals with an unknown arrival time, one constructs
a signal
estimator of the form

\begin{equation}
{\cal S}(t_0) ~=~ 4 Re \left[ \int_{0}^{\infty} df
{{\tilde h (f) \tilde h_s^* (f) e^{- i 2 \pi f t_0}} \over
{S_h(f)}} \right]
\label{eq:matchfilt}
\end{equation}

\noindent where $\tilde h (f)$ is the Fourier transform of the signal plus
noise,
$h(t)$, defined as

\begin{equation}
\tilde h (f) ~=~ \int_{-\infty}^{\infty} dt ~ h(t) e^{i 2 \pi f t}~\text{,}
\end{equation}
\noindent
$\tilde h_s (f)$ is the Fourier transform of the signal waveform and
$S_h(f)$ is the one-sided noise power spectral density.  An optimum
filter output is calculated for each realization of $h_s$ from Eq.
{\ref{eq:phasefunc}} using different values of the parameters.  The
approach of matched filtering thus requires the construction of a
``dense'' set of signal waveforms which cover the parameter space of
possible signals.

Other approaches to the variable frequency detection problem include
techniques based on either synthesizing a multichannel
filterbank\cite{jcp+97} or resampling the data at a variable
rate\cite{and93,smi87}. These techniques are used widely in the radio
pulsar community and, like the conventional matched filtering approach described
above, differ significantly from the algorithm presented here.

Just as the Fourier transform can be considered a form of matched
filtering using a dense set of sine and cosine functions appropriate
for periodic signals of unknown amplitude and phase, we wish to define
a transform which performs matched filtering for a dense set of
variable frequency signals with unknown parameters. By using the
Fourier transform for periodic signals, we avoid the need of
explicitly storing and computing the individual filter functions,
i.e. the sine and cosine functions in the time domain.  Analogously,
the appropriate transform for variable frequency signals will avoid
the need of generating a large set of filter functions and will
provide a prescription for densely covering the set of possible signal
waveforms.  We will informally call the transform for variable frequency
waveforms a ``chirp transform''.

The term ``chirp transform'' has been used elsewhere in the
literature.  For instance, Oppenheim et al. \cite{op99} describe a
``chirp transform algorithm'' (CTA) which is a special case of the
``chirp-z transform'' (CZT). The chirp-z tranform is well-known and
can be used to evaluate quadratic chirps.  The method described in our
paper is general and not constrained to quadratic chirp functions. We
call the algorithm described in this paper the ``fast chirp
transform'' (FCT).

The techniques discussed in this paper appear to be related to filter
bank design, to wavelet analysis, and to STFTs.  In fact, the chirp
transform can be viewed as a prescription for coherently adding the
outputs of a bank of variable-length STFTs with a particular
time-domain relationship.  Thus, the fast chirp transform may already
exist in some other formalism in the signal processing literature, but
we are currently unaware of it.  The recent work of Schutz
\cite{sch98} and Williams and Schutz\cite{will99} describes an
approach which in similar in several aspects, but differs in that
constant length STFTs are used rather than variable-length transforms.

In section {\ref{sec:matched}}, we give the discrete forms of the matched
filter outputs and discuss how these may be expressed in the form of the
discrete
analog to generalized Fourier integrals.  In section
{\ref{sec:chirptrans}} we derive the Two-Parameter Fast Chirp Transform
that can be used to
evaluate the discrete matched filter expressions. In section {\ref{sec:general}
we generalize the definition of the FCT to an arbitrary number of
parameters. In section
{\ref{sec:apps}} we briefly discuss several applications of the Fast Chirp
Transform.

\section{Discrete Matched Filtering}
\label{sec:matched}

The discrete forms of the time-domain and frequency domain matched
filter outputs (Eq. {\ref{eq:timematch}} and {\ref{eq:matchfilt}}) are
given by:

\begin{equation}
{\cal S}(\lambda_0,...,\lambda_M) ~=~ \sum_{j=0}^{N_0-1}
{h_s}(j,\lambda_0,...,\lambda_M)
h(j)\text{ , and}
\label{eq:disctimematch}
\end{equation}

\begin{equation}
{\cal S}(j_0,\lambda_0,...,\lambda_M) ~=~ {4 \over N_0} Re \left[
\sum_{k_0=0}^{N_0-1}
{{\tilde h(k_0) {\tilde h}_s^*(k_0,\lambda_0,...,\lambda_M) e^{-i 2 \pi j_0
k_0/N_0}} \over
{{S_h}(k_0)}} \right]
\label{eq:discfreqmatch}
\end{equation}
Appendix {\ref{app:stationary}} shows
that for variable frequency waveforms, $h_s(t)=A(t)
\cos(\phi(t,\lambda_0,...,\lambda_M))$, the discrete matched filter
outputs can be expressed as generalized Fourier integrals, and in
discrete notation take on a particularly simple form:

\begin{equation}
{\cal S}(\lambda_0,...,\lambda_M) ~=~
Re \left[\sum_{j_0 = 0}^{N_0-1} {\cal G}_{j_0} e^{ -
i\phi(j_0,\lambda_0,...,\lambda_M)}\right]
\label{eq:disctemp}
\end{equation}

\begin{equation}
{\cal S}(j_0,\lambda_0,...,\lambda_M) ~=~ {4 \over N_0} Re\left[\sum_{k_0 =
0}^{N_0-1}
{\tilde {\cal H}_{k_0}}  e^{- i \Phi(k_0,j_0,\lambda_0,...,\lambda_M)}\right]
\label{eq:discfreq}
\end{equation}

\noindent
where ${\cal G}_{j_0}$ and ${\tilde {\cal H}_{k_0}}$ can be
considered as the time or frequency series to be transformed,
and $\Phi(k_0,j_0,\lambda_0,...,\lambda_M)$ is a real phase function of the
form given in Eq. {\ref{eq:phi}}. 

In the next section we define the Chirp Transform and show how it can
be used to evaluate discrete transforms of the type shown
in Eq. {\ref{eq:disctemp}} and {\ref{eq:discfreq}}. Appropriate forms
of the FCT will replace both the forward and inverse transforms contained in
equations \ref{eq:disctemp} and \ref{eq:discfreq}. An inverse FCT is
not applicable to the detection problem and will not be considered in
this paper. The above formulation of the matched filtering process
included the starting phase as a search parameter. An alternative
approach is to convolve the signal with both the in-phase, cos($\phi$),
and quadrature-phase, sin($\phi$), filters and then sum the squares of
the results. This formulation is independent of the starting phase. In
order to make the chirp transform applicable to both formulations and
to a wider class of problems in general, the Re[] operator will not be
included in the definition. If necessary, this operator can be applied
after the transform.

\section{The 2-Parameter Fast Chirp Transform}
\label{sec:chirptrans}

\subsection{Example: The Quadratic Chirp (Linear Frequency Drift)}
\label{sec:quadchirp}
As an initial example, we consider the problem of the detection of a
quadratic chirp, e.g. a signal of the form $h_s(t) = A(t)\cos(2 \pi (f
t + {1 \over 2}\dot f t^2))$.  We wish to detect this
signal by matched filtering with a dense set of quadratic
chirp waveforms. This requires evaluation of sums of the form in Eq.
{\ref{eq:disctimematch}}. Note that if $\dot f$ were zero, the signal
waveform would be periodic and we might discretely sample the input
signal and then compute a power spectral estimate using the Fast
Fourier Transform (FFT).

Here we define a Fast Chirp Transform (FCT) for the quadratic chirp
analogous to the FFT.  This definition will be generalized to an arbitrary
parameter frequency waveform in the next section.  For simplicity we
first define the Discrete Chirp Transform (DCT) for the quadratic chirp in
analogy with the Discrete Fourier Transform (DFT):

\begin{equation}
H_{\{k_0,k_1\}} ~=~ \sum_{j_0=0}^{N_0 -1} h_{j_0} e^{i 2 \pi (k_0 (j_0/N_0)
+ k_1
(j_0/N_0)^2)}
\end{equation}

\noindent
where $h_{j_0}$ is the discretely sampled data.
The quadratic nature of the chirp is specified by the term $(j_0/N_0)^2$ in
the exponential. If $k_1$ is
zero, we have the usual DFT.

To derive the Fast Chirp Transform (FCT), we begin by breaking the
interval $\{j_0=0,...,N_0-1\}$ into a set of contiguous sub-intervals

\begin{equation}
H_{\{k_0,k_1\}} ~=~ \sum_{j_1=0}^{N_1-1}
~\sum_{j_0=j_0^{min}(j_1)}^{j_0^{min}(j_1+1) -1} h_{j_0} e^{i 2
\pi (k_0 (j_0/N_0) + k_1 (j_0/N_0)^2)}
\end{equation}

\noindent
where $j_0^{min}(j_1)$ is the lower boundary of each of the intervals.
The requirement that
the intervals be contiguous implies $j_0^{min}(j_1+1) = j_0^{min}(j_1)
+n_0(j_1)$,
where $n_0(j_1)$ is the number of points in the interval.

The boundaries are chosen as follows: we demand that the term $k_1 (j_0/N_0)^2$
change by no more than $\pi$ for values of $k_1$ appropriate for the
quadratic chirp signals being considered. If the term $k_1 (j_0/N_0)^2$
changes by
more than $\pi$ over a single sample, then the signal is not considered to
be finely enough sampled to resolve the phase evolution. This limitation is
analogous to the Nyquist sampling limit for periodic signals. Requiring
that the
term
$k_1 (j_0/N_0)^2$ remain relatively constant over a sub-interval allows us to
approximate the DCT as:

\begin{equation}
H_{\{k_0,k_1\}} \approx \sum_{j_1=0}^{N_1-1}
e^{i 2 \pi ( k_1 (j_0^{min}(j_1)/N_0)^2)}
\sum_{j_0=j_0^{min}(j_1)}^{j_0^{min}(j_1+1)
-1} h_{j_0} e^{i 2
\pi k_0 (j_0/N_0)}
\label{eq:conphase}
\end{equation}

\noindent
and we note that the second summation can now be computed as a
Fast Fourier Transform (FFT).  We further demand that the term
$k_1 (j_0^{min}(j_1)/N_0)^2$ increment by a constant amount from
one sub-interval to the next.
The requirement that the increment be less than $1/2$ for
the maximum value of $k_1$, $k_1^{max}$, specifies $N_1=2 k_1^{max}$,
with the ``Nyquist'' restriction that $k_1^{max} \le N_0/4$ for the
case of a quadratic chirp. We can then write:

\begin{equation}
H_{\{k_0,k_1\}} \approx \sum_{j_1=0}^{N_1-1}
e^{i 2 \pi  k_1 (j_1/N_1)}
\sum_{j_0=j_0^{min}(j_1)}^{j_0^{min}(j_1+1)
-1} h_{j_0} e^{i 2
\pi k_0 (j_0/N_0)}
\label{eq:fctquadc}
\end{equation}

\noindent
where we have anticipated that we will evaluate the
sum for integral values of $k_1$ using a standard FFT.
Finally, we express the inner sum in standard FFT form by
extracting a phase factor and we define the 2 parameter FCT,
$C_{\{k_0,k_1\}}$, as:

\begin{equation}
H_{\{k_0,k_1\}} \approx C_{\{k_0,k_1\}} \equiv \sum_{j_1=0}^{N_1-1}
e^{i 2 \pi  k_1 (j_1/N_1)}
\left[e^{i 2 \pi k_0 (j_0^{min}(j_1)/N_0)}
\sum_{j_0=0}^{n_0(j_1)-1} h_{j_0 + j_0^{min}(j_1)} e^{i 2
\pi k_0 (j_0/N_0)}\right]
\label{eq:fctquad}
\end{equation}

\noindent
As before, $n_0(j_1)$ is the number of points in the interval.  We
call $k_0$ and $k_1$ the ``conjugate variables'' of the linear and
quadratic terms, respectively, in the same sense that frequency and
time are a conjugate variable pair.  Note that both summations can be
implemented using FFTs and thus we are justified in calling the
transform a {\it fast} chirp transform.  The non-linearity of the
quadratic chirp is absorbed into the specification of the boundaries
of the contiguous intervals. This is the key concept of the FCT. Note
that the definition of $C_{\{k_0,k_1\}}$ is general and does not
depend explicitly on the quadratic nature of the phase evolution. The
same definition will apply to other non-linear phase evolution
functions.

We may also write:
\begin{equation}
C_{\{k_0,k_1\}} ~=~  \sum_{j_1=0}^{N_1-1}
e^{i 2 \pi  k_1 (j_1/N_1)} ~ C_{\{k_0,j_1\}}
\label{eq:fctquadb}
\end{equation}

\noindent
where we have used the notation $C_{\{k_0,j_1\}}$ to indicate that it is a
partial transform, i.e. transformed over one index, $j_0$, but not the
other, $j_1$. Equation {\ref{eq:fctquadb}} illustrates an interesting
feature of the FCT, which is important in implementation
considerations. Although the number of points in an interval,
$n_0(j_1)$, may be small, the partial transform, $C_{\{k_0,j_1\}}$,
requires evaluation at a large number of values of $k_0$, for example
at $N_0$ values of $k_0$.  This is equivalent to calculating the
oversampled FFT of the individual intervals with an oversampling
factor of $N_0/n_0(j_1)$. The problem of computing the
$C_{\{k_0,j_1\}}$ thus reduces to the problem of estimating the
oversampled spectrum of each interval. The oversampled spectrum may be
calculated exactly using the Fractional Fourier Transform (FRFT).
Fast methods have been developed for evaluating FRFTs \cite{bs91} and it
can be shown that the
calculation of the values of the individual oversampled FFTs that enter into
$C_{\{k_0,j_1\}}$ in Eqs. {\ref{eq:fctquad}} and {\ref{eq:fctquadb}}
requires $\text{O}(N_0 \text{log}_2 n_0(j_1))$ operations to
be computed exactly.  The computation required to calculate the entire
set of FRFTs can be shown to be $\text{O}(N_1 N_0 \text{log}_2 (N_0/N_1))$.
The phase factors, $\exp{(i 2 \pi k_0
(j_0^{min}(j_1)/N_0))}$ in Eq.  {\ref{eq:fctquad}} are easily computed
as part of the same formalism.

Taking into account the evaluation of the inner and outer
sums separately, the number of operations required for the evaluation of
the FCT can be shown to be:

\begin{equation}
N_{ops} \leq \text{O} (N_1 N_0 \text{log}_2 N_0 ) 
\label{eq:maxcomp}
\end{equation}

\noindent
which is at least as efficient as the matched filter approach.
The inequality indicates that the evaluation requires less computations
if approximations are used in evaluating the oversampled FFTs or if a
coarser sampling of the FCT over $k_0$ values is used.  
We will discuss the issue of computational efficiency further in Section
{\ref{sec:comp}} below.

\subsection{Accuracy of the Approximations}
\label{accapp}
How accurately can the FCT approximate the discrete matched filter
output?  There are three types of approximations to be considered:
(1) Possible use of the Stationary Phase Approximation in deriving a form for
the matched filter output; (2)
The constant phase approximation used to derive the discrete form
of the FCT; and (3)
Possible use of approximations in calculating the oversampled FFTs
that enter into the FCT.
We will consider each of these approximations in turn.

\medskip
\noindent
(1) There may be some error due
to the Stationary Phase Approximation itself (Appendix
{\ref{app:stationary}}) if
this is used to approximate the Fourier transform of the signal waveform.
This is not inherently part of the FCT, and we will
not discuss this approximation here since it depends on the specific
waveform being
analyzed. However, we note that the Stationary Phase Approximation can
be extremely accurate in practice \cite{dkp99,dis00}.
\medskip

\noindent
(2) The error due to the constant phase approximation (Eq.
{\ref{eq:conphase}}) in the FCT itself is obviously dependent on
the value of the conjugate variables at which the FCT is evaluated. If
the value of $k_1$ is small, the value of the increase in the
quadratic phase term as $j_1$ goes to $j_1 + 1$ is also small, and the
approximation will be very accurate.  By analogy to the Fourier
transform, we expect the ``frequency response'' of the output of the
matched filter computed by the FCT to behave similarly to the
frequency response of a power spectrum computed with an FFT.
Specifically, just as there is a roll-off in power for some periodic
signals near the Nyquist frequency, there will be a roll-off in the
accuracy of the FCT approximation to the matched filter output for
signals near the Nyquist limit of the conjugate variable, namely $k_1
(\text{Nyquist}) = N_1/2$. For values of $k_1$ well below Nyquist, we
expect the accuracy of the FCT approximation to the matched filter
output to be very good. Furthermore, if we wish higher accuracy in the
FCT approximation, we can employ the same techniques used in Fourier
analysis, namely, we can ``oversample'' the FCT. This can be
accomplished by ``zero-padding'' of the outer ($j_1$) FFT.

\medskip
\noindent
(3) Finally, there may be errors due to the possible use of approximations in
calculating the oversampled FFTs. As indicated in Eq. {\ref{eq:maxcomp}},
the exact calculation of the oversampled FFTs requires
$\text{O} (N_1 N_0 \text{log}_2 N_0 )$ calculations.  However, this assumes
that each interval is oversampled by a factor $N_0/n_0(j_1)$, which can
be quite large. Typically, oversampling factors of $2^2-2^3$ are sufficient
for an accurate approximation. If $N_1$ is large, the computational
requirements may thus
be reduced considerably by using such approximations
(i.e. $\text{O}(N_0 \text{log}_2 N_0)$
rather than
$\text{O}(N_1 N_0 \text{log}_2 N_0)$).
A quantitative discussion of oversampling approximations is beyond the scope
of this paper.

We also remark that the FCT formalism lends itself naturally to a
variety of hierarchical search approaches.  For instance, consider
Eqs.  {\ref{eq:fctquad}} and {\ref{eq:fctquadb}}.  The outer sum is a
coherent addition of the contributions from the individual
intervals. In order to implement a fast hierarchical search, we could
simply take the magnitudes of each $C_{\{k_0,j_1\}}$ and add them as an
incoherent sum over $j_1$, giving a measure of the incoherent power as
a function of $k_0$.  Values of $k_0$ with significant incoherent
power could then be examined in more detail by performing the coherent
summation over $j_1$.  This could lead to a dramatic decrease in the
required number of computations, depending on the threshold set for
the incoherent power summation step.

\subsection{Implementation Considerations}
\label{sec:implement}

As an initial trial, we have implemented the two-parameter FCT and tested
it on several
types of waveforms, including the quadratic chirp discussed above and
the ``Newtonian chirp'' discussed in Sec.\ref{sec:gw} below. We have
used two implementations, one which uses a fixed length oversampling
of the inner FFT, and one which uses a pre-packaged 2D FFT algorithm.

In this section we will describe in more detail the 2D FFT implementation
in order to give further insight into the details of the 2-parameter FCT
algorithm.
The 2D FFT implementation is extremely simple to code although it
is not the most computationally efficient. It will be
shown in the next section that even so, it is nearly as efficient as the
brute-force
matched filtering method.

The 2D FFT implementation of the FCT  packs the initial
one dimensional data array, $h_{j_0}$, into a sparse two dimensional
array $\hat{h}_{j_0,j_1}$. The packing will be determined by the
$j_0^{min}(j_1)$ array which is ultimately defined by the phase function
$\phi(j_0)$. Once the data are packed appropriately, the FCT is
calculated using any pre-packaged 2D FFT routine.
For the rest of this discussion, it is assumed that $N_0$ and
$N_1$ are compatible with the 2D FFT routine being employed. This
usually means that these lengths are a power of two.

From equation \ref{eq:fctquadc}, the two parameter FCT may be written as
\begin{equation}
C_{\{k_0,k_1\}} ~=~
\sum_{j_{1}=0}^{N_{1}-1} e^{i 2 \pi  k_{1} (j_{1}/N_{1})}
\sum_{j_0=j_0^{min}(j_1)}^{j_0^{min}(j_1+1) -1}
h_{j_0} e^{i 2 \pi k_0 (j_0/N_0)}.
\label{eq:2pfct}
\end{equation}
By defining a two dimensional $N_0 \times N_1$ array such that
\begin{equation}
\hat{h}_{j_0,j_1} = \left\{
		  \begin{array}{ll}
		   h_{j_0} & j_0^{min}(j_1) \leq j_0 < j_0^{min}(j_1 + 1) \\
		   0       & \mbox{otherwise}
		  \end{array}
		   \right.,
\label{eq:2dhj}
\end{equation}
equation \ref{eq:2pfct} becomes
\begin{equation}
\label{eq:2dfft}
C_{\{k_0,k_1\}} ~=~ \sum_{j_1 = 0}^{N_1 -1} \sum_{j_0 = 0}^{N_0 -1}
\hat{h}_{j_0,j_1} e^{i 2 \pi (k_1 j_1/N_1 + k_0 j_0/N_0)}
.
\end{equation}
As promised, $C_{\{k_0,k_1\}}$ is the 2D FFT of the sparsely packed
array $\hat{h}_{j_0,j_1}$. For the case of a monotonic phase function
that satisfies $\phi(0) = 0$ and $\phi(N_0) = N_1$, the array boundaries
are given by
\begin{equation}
j_0^{min}(j_1) = \phi^{-1}(j_1).
\end{equation}
Note that we have the freedom to rescale $\phi(j)$ so that $N_1$ may be
chosen arbitrarily.

The number of operations needed by the FCT implementation using a 2D
FFT is the same as that needed by a two dimensional FFT of order $N_0
\times N_1$:
\begin{equation}
N_{fct-2D} = \text{O}\left(N_0 N_1 \log_2{N_0N_1}\right).
\end{equation}

\subsection{Computational Efficiency}
\label{sec:comp}

For a matched filter operation using individual filters, there is
one Fourier transform to perform for each filter, thus the
number of computations, $N_{mf}$, is of order:
\begin{equation}
N_{mf} = \text{O}\left(N_1 N_0\text{log}_2 N_0\right)
\label{eq:matchedfilterorder}
\end{equation}
where $N_1$ is the number of filter functions needed to cover 
the space of possible waveforms, and $N_0$ is the number of samples in
the time series or frequency spectrum. As shown in section
\ref{sec:quadchirp} using the fractional fourier transform (FRFT), the
FCT computation can be of order
\begin{equation}
N_{fct-FRFT} \leq \text{O}\left(N_1N_0 \log_2 N_0\right).
\end{equation}

Two approaches offer potential computational
gains: (1) Reducing the oversampling factor in computing the inner FFT
(discussed in Section {\ref{sec:quadchirp}}), and (2) Relaxing the
requirement of full resolution in the $k_0$ variable.  Reducing the
oversampling factor can change the inner sum computation from an
$\text{O}\left(N_0 N_1 \log_2(N_0/N_1)\right)$ calculation to an
$\text{O}\left(N_0 \log_2{N_0}\right)$ calculation (see Section
{\ref{sec:quadchirp}}).  If $N_1 >> 1$ the total calculation of the
FCT will then be of order $\text{O}\left(N_0 N_1\text{log}_2
N_1\right)$. This is potentially a factor of
$\text{O}\left(\log_2{N_0}/\log_2{N_1}\right)$ more efficient than
the conventional matched filtering technique. We emphasize that the
detailed coefficients in front of the scalings are not yet known for
computationally efficient implementations.

Significant computational gains are also potentially available by
relaxing the requirement of full resolution in the $k_0$ variable.
Not all problems require the high $k_0$ resolution that the previously
discussed FCT implementations deliver. While the power is very
localized in the $k_0$ variable when the value of $k_1$ is that of
the actual signal(and vice-versa), the power can be significant for
values of $k_0$ and $k_1$ which simultaneously deviate from the actual
signal values.  The reason for this is that the deviation in $k_0$ can
be compensated for by a deviation in $k_1$, providing a reasonable
correlation of the matched filter template with the actual signal.
The determination of the optimum sampling resolution is closely related to
calculation of the {\it ambiguity function} \cite{hel68}.  Discussions
of techniques for calculating the ambiguity function have been given
by Owen \cite{Owe96}, Mohanty and Dhurandhar \cite{modhur96}, and
Owen and Sathyaprakash \cite{os99} for special cases of the types
of chirp functions considered here.

The FCT may be evaluated at lower resolution by reducing the order of the
2D array
$\hat{h}_{j_0,j_1}$ (Sec.  {\ref{sec:comp}}) from $N_0 \times N_1$ to
$(\epsilon N_0) \times N_1$ where $\epsilon < 1$. As long as
$(\epsilon N_0) \times N_1 > N_0$, we can still pack the original
$N_0$ data points into this smaller array. The resulting FCT will have
a coarser $k_0$ resolution but it will take $\text{O}(\epsilon N_1 N_0
\log_2{N_0})$ operations to perform (again neglecting terms of order
$\log_2{N_1}$).  Alternatively, in the straightforward evaluation of the
FCT contained in Eq. {\ref{eq:fctquadb}, the sums over $j_1$ are
simply evaluated at a decimated set of $k_0$.

We remark that sampling the FCT at lower resolution can be used as part of
a hierarchical algorithm, i.e. the FCT is first calculated at lower resolution
and values of $k_0$ are identified having excess signal strength.  The FCT is
then evaluated on a finer grid of $k_0$ values near those values of $k_0$
exhibiting the excess signal.

\subsection{Selection of the Range for Evaluation of the FCT}

We can select ranges of $k_0$ and $k_1$ to be evaluated by a process
analogous to a heterodyne operation on a periodic signal.  Selection
of the range of $k_0$ is done by the usual process of down-conversion
and low-pass filtering and we will not consider it further here.  For
the parameter $k_1$ the process is different in that no low-pass
filtering is required, and because the range of  $k_1$ can depend on $k_0$,
i.e. the conjugate parameter ranges need not be independent. This
provides considerable flexibility in determining the shape and volume
of parameter space that can be efficiently searched.

Suppose we wish to compute the FCT for a range of $k_1$ centered on
$k_1^{mid}$.
Re-writing the 2-Parameter FCT including the $k_1^{mid}$ term, we obtain:

\begin{equation}
H_{\{k_0,k_1\}} ~=~ \sum_{j_0=0}^{N_0 -1} h_{j_0} e^{i 2 \pi (k_0 (j_0/N_0)
+ (k_1+k_1^{mid} (k_0))
(j_0/N_0)^2)}
\end{equation}

\noindent
where we have explicitly allowed $k_1^{mid}$ to depend on $k_0$.
As before, assuming that $k_1 + k_1^{mid}$ is sufficiently small, we take the quadratic phase
term out of the inner summation to obtain:

\begin{equation}
H_{\{k_0,k_1\}} \approx C_{\{k_0,k_1\}} ~=~ \sum_{j_1=0}^{N_1-1}
e^{i 2 \pi  k_1 (j_1/N_1)}
\left[e^{i 2 \pi (k_0 (j_0^{min}(j_1)/N_0) + k_1^{mid} (k_0) (j_1/N_1))}
\sum_{j_0=0}^{n_0(j_1)-1} h_{j_0 + j_0^{min}(j_1)} e^{i 2
\pi k_0 (j_0/N_0)}\right]
\end{equation}

\noindent
To further simplify the notation for the outer sum, we define:

\begin{equation}
\label{eq:theta2d}
\Theta (k_0,j_1,j_0^{min},k_1^{mid}) = k_0 (j_0^{min}(j_1)/N_0) +
k_1^{mid} (k_0) (j_1/N_1)
\end{equation}

\noindent
to obtain

\begin{equation}
C_{\{k_0,k_1\}} ~=~ \sum_{j_1=0}^{N_1-1}
e^{i 2 \pi  k_1 (j_1/N_1)}
\left[e^{i 2 \pi \Theta (k_0,j_0^{min},k_1^{mid})}
\sum_{j_0=0}^{n_0(j_1)-1} h_{j_0 + j_0^{min}(j_1)} e^{i 2
\pi k_0 (j_0/N_0)}\right]
\label{eq:hetero2}
\end{equation}

\noindent
It is important to note that $N_1$ can be chosen to be
any integer less than or
equal to the ``Nyquist limit'', $N_0/2$.
This is due to the fact that we
can choose the boundaries, $j_0^{min}(j_1)$, of the intervals arbitrarily as
long as the phase change from the quadratic term is kept
sufficiently small.  Thus, we are free to choose any
range of $k_1$ around $k_1^{mid}$ as long as the
size of the range does not exceed the Nyquist limit.


\section{Generalized Definition of the FCT}
\label{sec:general}

Because the non-linearities of the phase evolution enter into the
FCT only in the boundaries of the sub-intervals, we may generalize the
definition of the FCT to the case of a sum of non-linear phase evolution
terms.  In particular, consider the DCT:

\begin{equation}
H_{\{k_0,..,k_{M-1}\}} ~=~ \sum_{j_0=0}^{N_0 -1} h_{j_0}
\text{exp}\left[{i 2 \pi \left(k_0 {j_0 \over N_0} + \sum_{p=1}^{M-1} k_p
{{\phi_p(j_0)} \over {\phi_p(N_0)}}\right)}\right]
\label{eq:generalchirp}
\end{equation}

\noindent
where the $\{\phi_p(j_0):p=1,...,M-1\}$ are a set of parameterless,
non-linear phase functions. The corresponding FCT is:
\samepage{
\begin{eqnarray}
\label{eq:generalchirpb}
C_{\{k_0,...,k_{M-1}\}} & ~=~
&\sum_{j_{M-1}=0}^{N_{M-1}-1} e^{i 2 \pi  k_{M-1} ({{j_{M-1}} \over
{N_{M-1}}})}...\\
& &~~ \times e^{ i 2 \pi k_1 ({{j_1^{min}(j_2,...j_{M-1})} \over {N_1}})}
\sum_{j_{1}=0}^{n_{1}(j_2,...,j_{M-1})-1}
e^{i 2 \pi  k_{1} ({{j_{1}}\over {N_{1}}})} \nonumber \\
& &~~ \times e^{ i 2 \pi k_0 ({{j_0^{min}(j_1,...j_{M-1})} \over {N_0}})}
\sum_{j_0=0}^{n_0(j_1,...,j_{M-1})-1}
e^{i 2 \pi k_0 ({{j_0} \over {N_0}})} h_{j_0+j_0^{min}(j_1,...j_{M-1})}
\nonumber
\end{eqnarray}
}

\noindent
where the $j_p^{min}(j_{p+1},...j_{M-1})$ and $N_p(j_1,...,j_{M-1})$
are specified by the phase functions $\{\phi_p(j_0):p=1,...,M-1\}$,
and the $\{N_p: p=1,...,M-1\}$ are determined by the maximum allowed
values of the $\{k_p:p=1,...,M-1\}$. As before, the intervals
specified by the $j_0^{min}(j_1,...j_{M-1})$ are contiguous. Thus for each
non-linear phase function $\phi_p(j_0)$ there is a corresponding
``conjugate variable'', $k_p$.


In analogy with equations \ref{eq:2dhj} and \ref{eq:2dfft}, the M
parameter FCT may be written in the form of an M-dimensional discrete
Fourier transform by defining the matrix $\hat{h}_{j_0,...,j_{M-1}}$ such that
\begin{equation}
\label{eq:generalh}
\hat{h}_{j_0,...,j_{M-1}} = \left\{ 
			\begin{array}{ll}
			h_{j_0} & j_p \leq \frac{\phi_p(j_0)N_p}{\phi_p(N_0)} < j_p + 1 \mbox{ for all $p \in [1,M-1]$}\\
			0 & \mbox{otherwise}
			\end{array}
			\right..
\end{equation}
With this definition, equation \ref{eq:generalchirpb} becomes:
\begin{equation}
\label{eq:ndfft}
C_{\{k_0,...,k_{M-1}\}} = \sum^{N_{M-1}-1}_{j_{M-1}=0} ... \sum^{N_{0}-1}_{j_{0}=0} e^{i2\pi\left(\frac{k_{M-1}j_{M-1}}{N_{M-1}}+...+\frac{k_{0}j_{0}}{N_{0}}\right)}\hat{h}_{j_0,...,j_{M-1}}.
\end{equation}
Note that the interval boundaries, $j_p^{min}$, may be determined from equation \ref{eq:generalh}.




Finally, we consider how the range for evaluation of the generalized FCT can be
specified, analagous to Eq. {\ref{eq:hetero2}} for the 2-parameter case.
To specify the region for which the FCT is to be evaluated, we 
add a term to each of the exponential terms in Eq. {\ref{eq:generalchirpb}}.
As for the 2-parameter case, equation \ref{eq:theta2d}, we define:
\begin{equation}
\Theta_p (k_{p-1},j_p,j_{p-1}^{min},k_p^{mid}) = k_{p-1} (j_{p-1}^{min}/N_{p-1}) + k_p^{mid}(j_p/N_p)
\end{equation}

\noindent
where $j_{p-1}^{min}$ is a function of $\{j_p,...,j_{M-1}\}$ and
$k_p^{mid}$ can depend on $\{k_0,k_1+k_1^{mid},...,k_{p-1} +
k_{p-1}^{mid}\}$. The parameter space searched by the FCT will then be
$k_p^{mid} \pm N_p/2$. This provides considerable flexibility
in determining the shape and volume of parameter space that can be
efficiently searched. We can now write:

\begin{eqnarray}
C_{\{k_0,...,k_{M-1}\}} & ~=~
&\sum_{j_{M-1}=0}^{N_{M-1}-1} e^{i 2 \pi  k_{M-1} ({{j_{M-1}}
\over {N_{M-1}}})}\\
& &~~ \times e^{ i 2 \pi \Theta_{M-1} (k_{M-2},j_{M-1},j_{M-2}^{min},k_{M-1}^{mid})}
\sum_{j_{M-2}=0}^{n_{M-2}(j_{M-1})-1}
e^{i 2 \pi k_{M-2} ({{j_{M-2}}\over {N_{M-2}}})} \nonumber \\
& &~~ \times ~ ...\\
\label{eq:heterochirp}
& &~~ \times e^{ i 2 \pi \Theta_{2} (k_1,j_2,j_1^{min},k_2^{mid})}
\sum_{j_{1}=0}^{n_{1}(j_2,...,j_{M-1})-1}
e^{i 2 \pi k_1 ({{j_{1}}\over {N_{1}}})} \nonumber \\
& &~~ \times e^{ i 2 \pi \Theta_{1} (k_0,j_1,j_0^{min},k_1^{mid})}
\sum_{j_0=0}^{n_0(j_1,...,j_{M-1})-1}
e^{i 2 \pi k_0 ({{j_0} \over {N_0}})} h_{j_0+j_0^{min}(j_1,...j_{M-1})}.
\nonumber
\end{eqnarray}
}



\section{Discussion: Application to Detection of Variable Frequency Signals}
\label{sec:apps}

\subsection{Detection of Gravitational Waves from the Binary Inspiral of
Neutron Stars and Black Holes}
\label{sec:gw}

One of the primary goals of the new generation of laser
interferometric gravitational wave detectors is the detection of
gravitational waves from the binary inspiral of compact objects,
specifically neutron stars (NSs) and black holes (BHs).  There is a
large literature written on the subject of matched filtering for
detection of gravitational waves using laser interferometers
(see e.g. \cite{th87,sch91,grasp}).
The matched filtering techniques
are based on Eq. {\ref{eq:matchfilt}} where $\tilde h(f)$ is the Fourier
transform of the gravitational strain generated from the differential
output of the interferometer, the $\tilde h_s (f)$ are the Fourier
transforms of theoretically generated binary inspiral signal
waveforms, and $S_h(f)$ is the measured power spectral density of
the interferometer.

A significant amount of work has gone into the calculation of binary
inspiral waveforms (called ``templates''), the spacing of such
templates to achieve near-optimal sensitivity, and the cost of
generating such templates in terms of compute cycles and storage
requirements\cite{apo95,Owe96,os99,mo98,tanaka00}.  Current matched filter
techniques require thousands to tens of thousands of templates to
cover the space of expected waveforms depending on the mass range of the
binary systems considered.

The method of chirp transforms described here does away with the
requirement of generating thousands of individual templates and
provides a natural way to cover the space of allowed
waveforms completely.  To apply the chirp transform to the binary inspiral
problem, we make use of the stationary phase formalism. Droz et
al. \cite{dkp99} have shown that the stationary phase formalism can be
used to provide an accurate approximation to the Fourier transform of
the time-domain waveforms of inspiraling binaries as calculated in the
``Newtonian'' approximation.  This is essentially an
application of the stationary phase approximation discussed in
Appendix {\ref{app:stationary}} to the case of gravitational
waveforms. Damour et al. \cite{dis98,dis00} have shown that the binary
inspiral waveforms can be accurately calculated using the SPA and an
alternative formalism based on ``P-approximants''. They note that care must
be taken in the treatment of the termination of the waveform at the time
of the final plunge and merger.

In order to illustrate the use of the FCT in gravitational wave
detection, we discuss the example of a ``post-Newtonian'' (PN) expansion.
The canonical PN stationary phase waveforms for binary
inspiral up to 2PN order are of the form\cite{os99}:
\begin{equation}
\tilde h_s (f)~=~\left( {{5 \mu} \over {96 \Msol}}\right)^{1/2}
\left( M_{tot} \over {\pi^2 \Msol}\right)^{1/3}
f^{-7/6} \Tsol^{-1/6} exp[i \Psi (f) ]
\label{eq:gwform}
\end{equation}

\begin{eqnarray*}
\Psi (f) & ~=~ & 2 \pi f t_c - 2 \phi_c - \pi/4 \\
& & \mbox + ~ {3 \over {128 \eta}} \left[ \chi^{-5/3} + \right.
\left({3715 \over 756} + {55 \over 9} \eta \right) \chi^{-1} - 16 \pi
\chi^{-2/3} \\
& & \mbox + ~ \left.\left( {15293365 \over 508032} + {27145 \over 504} \eta +
{3085 \over 72} \eta^2
\right) \chi^{-1/3} \right]
\end{eqnarray*}

\medskip
\noindent
where, for simplicity, we consider only one polarization.  The variables have
been
defined as usual ($\Msol$ is the mass of the sun, $\Tsol$ is $G \Msol/c^3$ and
has a value of approximately $4.925\times 10^{-6} \text{sec}$, $M_{tot}$ is
the total mass of the binary system, $\eta = \mu
/ M_{tot}$, $\mu$ is the reduced mass of the binary, $t_c$ is the time of
coalescence, $\phi_c$ is the phase at coalescence), and we have defined
\begin{equation}
\chi ~=~ {{\pi M_{tot} \Tsol} \over \Msol}f
\end{equation}

It can be seen that the stationary phase waveforms have frequency dependent
amplitudes and phase functions that are expansions in powers of the
frequency, $f$. In particular,

\begin{equation}
\Psi (f)  ~=~ \alpha + 2 \pi t_c  f + \lambda_0 f^{-5/3} + \lambda_1 f^{-1} +
\lambda_{1.5} f^{-2/3} + \lambda_2 f^{-1/3}.
\end{equation}

\noindent
where $\alpha$ is a phase constant and the $\lambda_x$ are
coefficients of the frequency expansion which depend on $M_{tot}$,
$\Msol$, $\Tsol$, and $\eta$.

In order to apply the FCT, we construct the discrete version of the matched
filter
output, Eq. {\ref{eq:discfreqmatch}}, where $\tilde h (k_0)$ is the Fourier
transform of the discretely sampled interferometer strain output,
${\tilde h}_{s} (k_0)$ are the stationary phase waveforms given in
Eq. {\ref{eq:gwform}} above, and $S_h (k_0)$ is the noise power spectral
density of the interferometer.

The FCT is then used to evaluate the matched filter, with a resulting
output,

\begin{equation}
C_{\{t_{c},\lambda_0,\lambda_1,\lambda_{1.5},\lambda_2\}}
\end{equation}

\noindent
In this expression, $2 \pi t_c$ is the conjugate variable of the
linear $f$ term; $\lambda_0$ is the conjugate variable to the
Newtonian term, $f^{-5/3}$, $\lambda_1$ is the conjugate variable to
the 1PN term, $f^{-1}$, etc. Numerous considerations arise in
selecting the search ranges for the conjugate parameters.  As
discussed by several authors \cite{mo98,os99}, spinless low-mass
binaries should be reasonably well detected by a three parameter
search over $\{t_{c},\lambda_0,\lambda_1\}$.  The conjugate parameters
$\lambda_0$ and $\lambda_1$ fulfill a function similar to the
parameters $\tau_0$ and $\tau_1$ that appear in the literature
(e.g. \cite{mo98,os99}) which represent the Newtonian and 1PN
contributions to the time to coalescence, respectively.  Thus, in the
case of spin-less low-mass binaries, $\lambda_{\{1.5,2\}}$ can be
considered functions of $\lambda_{\{0,1\}}$. Owen and Sathyaprakash
\cite{os99} point out that $\tau_{\{0,1.5\}}$ may be more convenient
search parameters. Hence, $\lambda_{\{1,2\}}$ become functions of
$\lambda_{\{0,1.5\}}$.  In this case, the ``heterodyne'' approach (see
the discussion following Eq.  {\ref{eq:ndfft}}) can be used with
the dependent parameters to reduce the search space to that of
spin-less 2PN waveforms.

To search for binaries with spin, additional independent parameters will
be needed and thus it will be useful to search in a range
around the spin-less PN expansion (or other expansion). This can also be
accomplished using
the method described in the discussion following Eq. {\ref{eq:ndfft}}.
The technique will be particularly useful for massive binaries for which spin
interactions could be significant.  An important step will be to estimate
limits to the range of the conjugate variables in the FCT analysis due to
spin effects.  The FCT then provides a formalism for searching the
complete waveform space, even if the exact waveforms are not known.

It will also be quite useful to enlarge the search region beyond the
space physically accessible by astrophysical binary systems. While no
binaries are expected outside the physically accessible regions, it is
important to study the characteristics of noise signals in regions
close to the physically accessible regions.  The FCT formalism
provides a straightforward way to tailor the analysis to a range of
search regions. 
This, of course, is also possible with conventional
template-based techniques.

The FCT formalism may be useful for expansions other than
post-Newtonian. In particular, we are very interested to see whether
the FCT formalism can be applied to the P-approximants discussed very
recently by Damour et al. \cite{dis00}. Also, as we remarked earlier,
the FCT lends itself naturally to hierarchical approaches for 
binary inspiral detection.  We note in particular the recent paper
by Tanaka and Tagoshi\cite{tanaka00} which discusses efficient
hierarchical search algorithms which have several similarities to
the general FCT algorithm.

In summary, the use of the FCT for detection of the chirps from
gravitational waves has several attractive features.  First, no
explicit calculation and storage of gravitational waveforms is
required for the analysis.  Only the order of the PN expansion,the
power-law exponents appearing in the expansion, and the range of the
search parameters is important.  Second, waveforms with perturbations
on the phase evolution such as those due to spins can be detected even
if the exact waveforms are not known since the FCT can be used to
search completely an arbitrary region of parameter space.  The only
requirement is that the perturbation not involve significant terms
beyond those in the expansion considered for the FCT.  The FCT may
therefore be very useful in the search for BH-BH coalescence where the
waveforms are not precisely known \cite{bct98}, or for sources to be
detected by the space-based gravitational detector, LISA, where the
waveforms are also only approximately known and phase perturbations
are likely to be present. Finally, the FCT formalism can be used to
investigate the characteristics of noise signals in the neighborhood
of expected signals from binary inspiral.

\subsection{Detection of Rotating Neutron Stars}
\subsubsection{Acceleration Searches}
Pulsars are rotating neutron stars that spin with periods in the range
of approximately 1 ms up to hundreds of seconds.  Pulsars are detected
primarily at radio and x/$\gamma$-ray wavelengths.  In the future,
rotating neutron stars may also be detectable as sources of
gravitational waves. Detection of pulsars usually employs Fourier
transform techniques to find the periodic pulses.  However, several
effects complicate the detection of pulsars and cause the pulsations
to deviate from being strictly periodic. For instance, the emission
from pulsars in compact binary systems is Doppler shifted causing a
frequency variation on the time scale of the orbital period.
Likewise, the earth's rotation and orbit can induce frequency and
phase variations that are dependent on the position of the source on
the sky.  Rotating neutron stars can also have non-negligible spin
down effects, especially if the neutron star is young.  Any of these
effects can be important at both radio and x/$\gamma$-ray wavelengths
depending on the length of the observation. They are also likely to be
important in future searches for gravitational wave emission from
rotating neutron stars due to r-modes, or from older rotating neutron
stars because of the earth's orbit and rotation.

In the past, so-called ``acceleration searches'' have been employed to
detect pulsars with slowly varying frequency \cite{and93}. These are
essentially matched filter techniques implemented either with
templates, or equivalently, with ``stretching'' of the time or
frequency variable. This requires individual matched filter operation,
one for each discrete acceleration trial. The FCT analog is that of
the quadratic chirp analysis discussed in Sec.
{\ref{sec:chirptrans}}. The FCT also provides a natural extension to
searches beyond quadratic (acceleration) effects.

\subsubsection{Dispersion Measure Searches}
Radio radiation emitted by pulsars travels through a diffuse
interstellar plasma known as the interstellar medium (ISM) before
reaching detectors on Earth. The dispersive properties of the ISM
cause individual radio pulses to broaden in time. This dispersion
broadening will reduce the chances of detecting a given pulsar
signal. The magnitude of the dispersion effect is measured by a
quantity called the dispersion measure (DM). If the DM is known, the
dispersion effect can be removed from the pulsar signal using standard
digital signal processing (DSP) techniques. When searching for new
pulsars, the DM is rarely known and systematic searches must be
performed both in DM and in pulsar period.

The effect of ISM dispersion may be removed from the received signal by
applying the following transformation \cite{hr75,jcp+97}:
\begin{equation}
S(DM,t) = \int_{-\infty}^{\infty}\hat{S}_r(f)e^{i2\pi f t}e^{i 2\pi DM
\phi(f)} df
\end{equation}
where $\hat{S}_r(f)$ is the Fourier transform of the received signal
and $S(DM,t)$ is known as the dedispersed signal. When searching for
new pulsars, one must generate several time series corresponding to
different values of the DM. The above equation shows that $S(DM,t)$ is
simply a chirp transform of $\hat{S}_r(f)$. The FCT provides an
efficient way to generate $S(DM,t)$ for several values of DM. Each of
these time series can then be searched for periodic signals.

When searching for pulsar signals, one typically ``detects'' the total
power in the signal by calculating $P(DM,t) = \|S(DM,t)\|^2$ and then
averages over a small window of time. Each time series is then
searched separately using various pulsar detection techniques. The
structure of the FCT points to the possibility of a slightly different
technique. Rather than searching each time series separately, one
first calculates $P(t) = \sum_{DM} P(DM,t)$ and then searches this
time series for possible pulsar signals (see section \ref{accapp}). Using the
property that the sum of the squares is conserved under a Fourier
transform, the second set of FFTs in the FCT does not need to be
performed in order to calculate $P(t)$. Thus, a highly efficient
intermediary chirp transform can be used instead of the complete
FCT.


\section{Summary}
We have described an algorithm for the detection of signals with variable
frequency. Standard detection algorithms use matched
filtering techniques which require both the computation of a large set
of task specific filter functions and a prescription for densely
covering the set of possible signal waveforms. The Fast Chirp
Transform proposed in this paper automatically precludes the need to
generate specific filter functions since standard FFTs can be used in
the implementation and the FCT immediately provides the prescription
for densely covering the waveform parameter space.

The FCT for a two parameter chirp was defined and then
generalized to N parameters with arbitrary phase functions. A
straight forward implementation of the FCT was discussed and it was
shown to be comparable in efficiency with  the brute-force matched filtering
approach. Several approaches to achieving even better computational efficiency
were also discussed.

The efficient detection of variable frequency signals has a large
number of practical applications. Of considerable interest to the
authors is the detection of gravitational waves from NS and BH binary
systems and the detection of radio waves from pulsars. Another obvious
area of application is radar-sonar signal processing where target or
transmitter motion can cause Doppler frequency shifts in the received
signal.  Other potential areas of application include communications
and image processing.
A more detailed description of the FCT and its
application to the above problems will be the subjects of future work.

We acknowledge informative and pivotal discussions with Ben Owen and
B. Sathyaprakash. We also acknowledge helpful and stimulating
discussions with Alessandra Buonanno, Tibault Damour, Scott Hughes and
Albert Lazzarini.  This work was supported in part by the LIGO
Laboratory under cooperative agreement NSF-PHY-9210038 and by grant NSF-PHY-9970877. This paper is LIGO document LIGO-P000003-00-R.

\appendix
\section{The Stationary Phase Approximation and Matched Filtering}
\label{app:stationary}

\subsection{Frequency-Domain Matched Filtering}

We begin by showing how the Fourier transform of waveforms such as
those of Eq.  {\ref{eq:phasefunc}} can be approximated in a way that
allows them to be expressed naturally in frequency-domain matched
filtering, Eq. {\ref{eq:matchfilt}}.  The Stationary Phase
Approximation (SPA) (see e.g. \cite{bo78}; see also Ref. \cite{th87}
and Refs. \cite{finnchern93,dis00} for a description in the context of
gravitational wave detection), provides a prescription for
approximating the Fourier transform of a function of the form
$h_s(t)=A(t)\cos{(\phi(t))}$ (where $A(t)$ and $\phi(t)$ are real and
$\phi '(t)$ is positive):

\begin{equation}
\tilde h_s (f) ~=~ {1 \over 2} \left[
\int_{-\infty}^{\infty} dt~A(t)~e^{i f\psi_+(t)}
+ \int_{-\infty}^{\infty} dt~A(t)~e^{-i f\psi_-(t)}\right]
\label{eq:spasplit}
\end{equation}

\noindent
with $\psi_\pm(t) = \phi(t)/f \pm 2 \pi t$. If $t_f$ exists such that
$\psi_+'(t_f) = 0$ or $\psi_-'(t_f) = 0$, then $t_f$ is called a
``stationary point''. Considering positive $f$ and positive $\phi
'(t)$, only the second integral in Eq. {\ref{eq:spasplit}} contains a
stationary point. Hence, to leading order, we can write \cite{bo78}:
\begin{equation}
\tilde h_s (f) ~\sim~ {1 \over 2}
\int_{-\infty}^{\infty} dt~A(t)~e^{i ( 2 \pi f t - \phi(t))}
\end{equation}
for $f>0$. Note that we compute $\tilde h_s(-f)$ using the Fourier
transform property of real functions: $\tilde h_s(-f) = \tilde
h_s^*(f)$.

If all derivatives of $\psi_-(t)$ up to order $p$ are zero
at $t_f$, then the Fourier transform of $h_s$ may be approximated by:
\begin{equation}
\tilde h_s (f) ~\sim~ {\cal A}(f) \text{exp}\left[ i \Psi (f) \right]~,
\label{eq:statphase}
\end{equation}
\noindent
with components given by somewhat complicated but straightforward
expressions: \begin{equation}
{\cal A}(f) ~=~ A(t_f) \left[ {{ p!} \over {f | \psi^{(p)}(t_f)|}}
\right]^{1/p}
{{\Gamma (1/p)} \over p}~,\text{ and}
\end{equation}
\begin{equation}
\Psi (f) ~=~ 2 \pi f t_f - \phi(t_f) \pm \pi/2 p
\end{equation}
\noindent
where the sign of $\pi/2 p$ is positive or negative depending on
whether $\psi^{(p)}_-(t_f)$ is positive or negative, respectively.
In particular, for $p=2$, the following approximation holds:

\begin{equation}
\tilde h_s(f) ~=~ \sqrt{\pi \over 2} A(t_f) |\phi''(t_f)|^{-1/2} e^{i ( 2
\pi f t_f
- \phi(t_f) \pm
\pi/4)}.
\end{equation}
The stationary phase approximation is accurate as long as the amplitude of
$h_s$
does not vary too quickly compared to the time derivative of the phase,
$\phi '(t)$, and the effect of the higher derivatives of $\phi(t)$ on the
phase evolution are small compared to the effect of $\phi '(t)$.

Using the form given in Eq. {\ref{eq:statphase}}, we can now rewrite
Eq. {\ref{eq:matchfilt}}.
Gathering all the amplitude terms together,

\begin{equation}
\tilde {\cal H}(f) ~=~ {{{\tilde h (f)} {{\cal A}^*(f)}} \over {S_h(f)}
}\text{ ,}
\end{equation}

\noindent
we can express the matched filter output as:

\begin{equation}
{\cal S}(t_0) ~=~ 4~ \text{Re} \left[ \int_{0}^{\infty} df
{\tilde {\cal H}(f)} e^{- i \Phi (f)} \right]
\end{equation}

\noindent
where
\begin{equation}
\Phi(f) = \Psi(f) - 2 \pi f t_0.
\label{eq:phi}
\end{equation}
Hence, the matched filtering operation in the frequency
domain is expressed as an integral transform,
specifically a so-called generalized
Fourier integral.  In analogy
with the Discrete Fourier Transform (DFT), we can write this in discrete
form as:

\begin{equation}
{\cal S} ~=~ {4 \over N_0} \text{Re} \left[ \sum_{k_0 = 0}^{N_0-1}
{\tilde {\cal H}_{k_0}}  e^{- i \Phi_{k_0}} \right]
\label{eq:disctrans}
\end{equation}

\noindent
The summation can be computed as a FCT.

\subsection{Time-Domain Matched Filtering}

We note that for signal waveforms of the form,
$h_s(t)=A(t)\cos{(\phi(t))}$, the expression (Eq. {\ref{eq:timematch}})
for time-domain matched filtering yields directly:

\begin{equation}
{\cal S} ~=~
\text{Re} \left[ \int_{0}^{T} dt~ h(t) A(t) e^{ i \phi(t)} \right]
\label{eq:timeint}
\end{equation}

\noindent
Such  signals are of considerable interest and include
periodic signals with frequency drift.
The integral transform in Eq. {\ref{eq:timeint}}
can be represented in discrete form in the usual way as:

\begin{equation}
{\cal S} ~=~
\text{Re} \left[ \sum_{j_0 = 0}^{N_0-1} {\cal G}_{j_0} e^{ i\phi(j_0)}\right]
\label{eq:disctime}
\end{equation}

\noindent
where
\begin{equation}
{\cal G}_{j_0} = h(j_0)A(j_0)
\end{equation}

\noindent
and where $h(j_0)$, $A(j_0)$, and $\phi(j_0)$ are the discretely
sampled values of the continuous functions.


\end{document}